# EPIHC: Improving Enhancer-Promoter Interaction Prediction by using Hybrid features and Communicative learning

Shuai Liu, Xinran Xu, Zhihao Yang, Xiaohan Zhao, Wen Zhang*

*Corresponding author: Wen Zhang, College of Informatics, Huazhong Agricultural University, Wuhan 430070, P. R. China. E-mail: zhangwen@mail.hzau.edu.cn (or) zhangwen@whu.edu.cn

**Abstract**

Enhancer-promoter interactions (EPIs) regulate the expression of specific genes in cells, and EPIs are important for understanding gene regulation, cell differentiation and disease mechanisms. EPI identification through the wet experiments is costly and time-consuming, and computational methods are in demand. In this paper, we propose a deep neural network-based method EPIHC based on sequence-derived features and genomic features for the EPI prediction. EPIHC extracts features from enhancer and promoter sequences respectively using convolutional neural networks (CNN), and then design a communicative learning module to captures the communicative information between enhancer and promoter sequences. EPIHC also take the genomic features of enhancers and promoters into account. At last, EPIHC combines sequence-derived features and genomic features to predict EPIs. The computational experiments show that EPIHC outperforms the existing state-of-the-art EPI prediction methods on the benchmark datasets and chromosome-split datasets, and the study reveal that the communicative learning module can bring explicit information about EPIs, which is ignore by CNN. Moreover, we consider two strategies to improve performances of EPIHC in the cross-cell line prediction, and experimental results show that EPIHC constructed on training cell lines exhibit improved performances for the other cell lines.

**Availability:** The codes and data are available at example@

## Introduction

Enhancers and promoters are two types of important DNA sequences in cells. An enhancer is about 50-1500 base pairs (bp) in length and has important roles in controlling the transcription of specific genes[1]. The length of a promoter ranges from 100-1000 bp, and determines the starting point position of gene transcription[2]. Recent studies on the three-dimensional genome [3] show that the distal enhancers can interact with the proximal promoters to regulate the expression of target genes. Enhancer-Promoter Interactions (EPIs) are important for understanding gene regulation, cell differentiation and disease mechanisms[4]. For example. that mutations in enhancers and promoters, which lead to the changes in EPIs, are responsible for diseases such as $\beta$-thalassemia and congenital heart disease[4, 5].

EPI identification through wet experiments is costly and time-consuming, and computational methods are in demand. A great number of computational EPI prediction methods have been proposed, and are roughly categorized as machine learning-based methods and deep learning-based methods.

The machine learning-based methods make use of the genomic or sequence-derived features of enhancers and promoters, and then adopt the classification methods to build prediction models. The earlier machine learning-based methods use the genomic features, such as the measures of DNase-seq, DNA methylation, histone markers, transcription factors, ChIP-seq, etc.. For example, IM-PET[7] used the structure, function and evolution features of enhancers and promoters, and built the random forest (RF)-based EPI prediction model; RIPPLE[8] combines Random Forest and group LASSO to select features from histone marks and transcription factors of the enhancer-promoter pairs, and then trains the RF-based EPI predictor; TargetFinder[9] utilizes Gradient Boosted Regression Trees (GBRT) model with functional genomic signatures of enhancer-promoter pair to build the prediction model, and found that the genomic window between enhancer-promoter pairs is important for the EPI prediction. More recently, the machine learning-based methods combine genomic and sequence-derived features. For example, PEP[10] takes advantage of transcription factors binding site (TFBS) motifs to form the PEP-motif module and uses DNA sequence embedding to form the PEP-word module, and then trains gradient tree boosting classifier on features from these two modules for the EPI prediction; EP2vec[11] applies the Doc2vec model to obtain the DNA sequence representations and integrates genomic features into the GBRT classifier to predict EPIs.

With the rapid development of deep learning, recurrent neural network (RNN) and convolutional neural network (CNN) have been applied to the EPI prediction. For example, EPIANN[12] encodes the enhancer and promoter sequences with one-hot encoding, and utilizes CNN to extract embeddings, and finally uses a fully connected network for EPI prediction; SPEID[13] combines CNN and Long Short-Term Memory (LSTM) network, and takes the long-term dependency in DNA sequences into account; SIMCNN[14] utilizes CNN with lots of kernels to build the prediction models. EPIVAN[15] employs pre-trained DNA2vec vectors to obtain sequence embedding, and then applies CNN and RNN to learn the representation of enhancers and promoters, and then uses the attention mechanism to aggregate features for the EPI prediction.

Although great efforts have been made on the EPI prediction, there still exists room for improving performances. First, the deep learning techniques have shown great potential in the EPI prediction. Recurrent Neural Network (RNN), which shares weights across timesteps and leverages historical information to capture segment-level dependency in enhancer and promoter sequences efficiently, is often used after CNN in EPI prediction models[13, 15], but it needs lots of training time. Second, existing deep learning-based methods directly concatenate the learned representations of enhancer and promoter sequences, but oversight the communicative information between enhancers and promoters. Third, most existing EPI prediction methods only pay attention to exploit either genomic features or sequence features, rarely combine them for EPI prediction. Fourth, most of the existing EPI prediction methods are cell line-specific, whereas they perform unsatisfyingly in the cross-cell line prediction[14, 16].

To address the above issues, we propose a neural network-based EPI prediction method called EPIHC by using hybrid features and communicative learning. The highlights of EPIHC are as follows:

(1) We design a communicative learning module, which captures segment-level communicative information between enhancer and promoter sequences as well as the sequence dependency. More importantly, the module has good computational efficiency.

(2) EPIHC utilizes hybrid features, including both sequence-derived features and genomic features, and the combination of diverse information can lead to high-accuracy prediction models. EPIHC produces better performances than other state-of-the-art methods on both benchmark datasets and chromosome-split datasets.

(3) We propose two strategies including data-ensemble and model-ensemble to enhance the performances of EPIHC in the cross-cell line prediction. Experimental results show that models trained on available cross-cell lines perform well on other cell lines.

## Materials and Methods

### Datasets

The EPI benchmark datasets used in the study come from TargetFinder[9], and have been used by SPEID[13, 17], SIMCNN[14] and EPIVAN[15]. The details of the datasets are shown in Table 1. The datasets consist of EPI data from six groups of human cell lines, including lymphoblasts (GM12878), umbilical vein endothelial cells (HUVEC), epidermal keratinocytes (NHEK), ectodermal lineage cells of cervical cancer patients (HeLa-S3), fetal lung fibroblasts (IMR90) and mesoderm lineage cells (K562) of leukemia patients. For each cell line, the confirmed enhancer-promoter interactions are used as positive samples, and 20 negative samples are randomly sampled per positive sample from non-interacting pairs[9, 18]. EPI datasets contains the coordinates of promoter-enhancer pairs on genomes from which the sequences can be obtained, and the genomic data, such as transcription factors and histone markers, are also available.

**Table 1.** Statistics of EPI benchmark datasets

| Cell lines | Positive samples | Negative samples |
| --- | --- | --- |
| GM12878 | 2113 | 42200 |
| HUVEC | 1524 | 30400 |
| HeLa-S3 | 1740 | 34800 |
| IMR90 | 1254 | 25000 |
| K562 | 1977 | 39500 |
| NHEK | 1291 | 25600 |

### Architecture of EPIHC

As shown in Figure 1(a), EPIHC extracts sequence-derived features about enhancer and promoter sequences using CNN, and then learn the communicative features between enhancers and promoters through a communicative learning module; EPIHC also takes genomic features of enhancers and promoters into account. Then, EPIHC combines the sequence-derived features and genomic features to build the EPI prediction model.

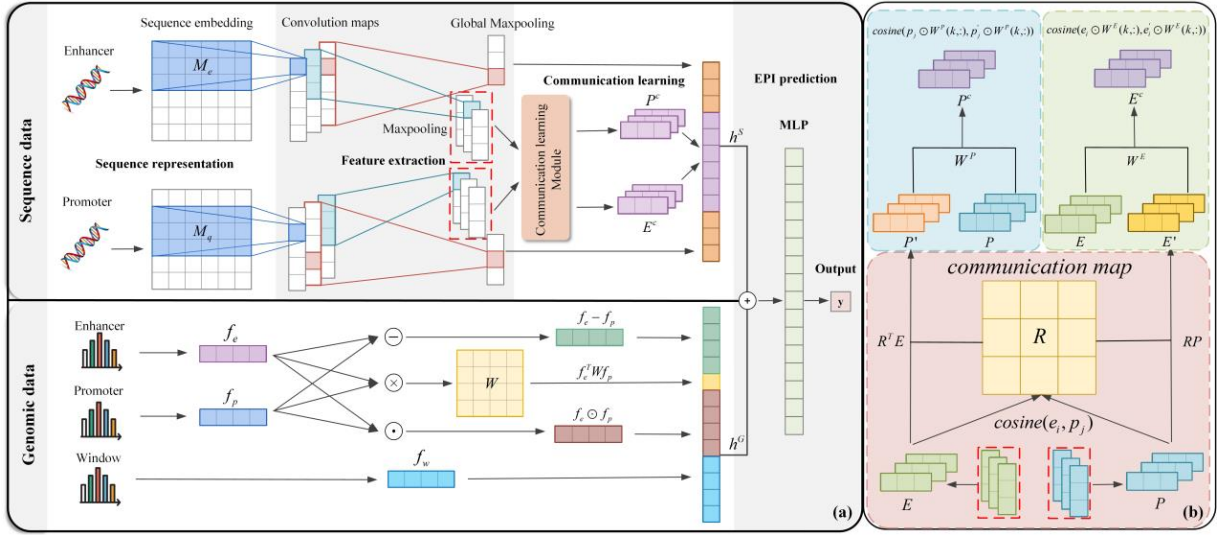

**Figure 1.** The overview of the architecture of EPIHC. (a) shows the enhancer-promoter pairs pass through the sequence-derived feature learning module and genomic feature learning module, then features are combined to make the prediction. "-", "×", "·", "+" mean the operators of subtraction, bilinear similarity, Hadamard product and concatenate between two vectors, respectively; (b) shows the design of the communicative learning module.

### Sequence-derived feature learning

*Sequence representation:* First, we follow the approach in [13] to expand the lengths of enhancer and promoter sequences to 3000 and 2000 respectively by extending the original sequences with flanking nucleotides. Second, we obtain $L-k+1$ kmers from an enhancer/promoter sequence with $L$ nucleotides as [15]. Third, we use pre-trained vectors calculated by DNA2vec[19] to represent kmers. DNA2vec[19] is a DNA representation learning algorithm based on Word2vec[20] framework, and then calculate the pre-trained vectors (dimension is set to 100) based on the entire human genome. At last, we obtain an embedding matrix with size $(L-k+1) \times 100$ for an enhancer/promoter sequence, whose rows are the pre-trained vectors of the corresponding kmers.

*Feature extraction:* We apply a convolution with size $s_{ck} \times 100$ and kernel number $n_f$ to the embedding matrix of each enhancer/promoter sequence[21]. Then, we respectively adopt global max pooling and max pooling with size $s_{mp} \times 1$ to obtain one feature map with size $n_f \times 1$ and $n_f$ feature maps with size $N \times 1$, where $N = \left\lfloor \frac{(L-k+1)-s_{ck}+1}{s_{mp}} \right\rfloor$. The above feature maps of the enhancer and the promoter obtained by max pooling are concentrated by column to form two feature matrices $E(N^E \times n_f)$ and $P(N^p \times n_f)$. Each row of feature matrices responds to a specified segment of sequences, and we take the row vectors as the contextual vectors for the enhancers and promoter, denoted as $e_i$ ($1 \times n_f$) and $p_j$ ($1 \times n_f$), $i = 1,2,\cdots,N^E, j = 1,2,\cdots,N^p$.

*Communicative learning:* The communications between segments of an enhancer and a promoter can bring explicit interactions between the enhancer and promoter. As shown in Figure 1(b), inspired by [22], we design a communicative mechanism to learn the communicative features of enhancers and promoters. The basic idea of communicative learning is to compare each contextual vector of a promotor (or enhancer) against all contextual vectors of an enhancer (or promoter). The pairwise communication between their segments are modeled in a communication map with size $N^E \times N^p$, which is computed according to the following equation:

$$R = (\alpha_{i,j}), i = 1,2,\cdots,N^E, j = 1,2,\cdots,N^p.$$

where $\alpha_{i,j} = cosine(e_i, p_j)$ is the similarity score between contextual vectors $e_i$ and $p_j$. The communication map $R$ is row-normalized with row sum equals to 1 for the next step.

Enhancer contextual vectors' influence on promoter contextual vectors and vice-versa is computed according to the following equations:

$$E' = RP$$
$$P' = R^T E$$

where $E'$ ($N^E \times n_f$) and $P'$ ($N^p \times n_f$) are the promoter features and enhancer features weighted by their contribution to the enhancer-promoter communication.

We design the multi-context communication function to compute the communicative features:

$$E^c(i,k) = \text{cosine}(e_i \odot W^E(k,:), e'_i \odot W^E(k,:))$$
$$P^c(j,k) = \text{cosine}(p_j \odot W^P(k,:), p'_j \odot W^P(k,:))$$
$$i = 1,2,\cdots,N^E, \ j = 1,2,\cdots,N^p, \ k = 1,2,\cdots,K$$

where $e'_i$ and $p'_j$ are row vectors of $E'$ and $P'$, and $W^E$ and $W^P$ are trainable matrices with the size $N^E \times n_f$ and $N^f \times n_f$, and $\odot$ is Hadamard product. $E^c$ and $P^c$ are the communicative features with size of $N \times K$, and their rows of them represents the communicative values from the $K$ perspectives.

Finally, we adopt the max value of each row of $E^c$ and $P^c$, and combine them with the global max pooling features to generate $h^S$, which represents the sequence-derived features of the enhancer-promoter pairs.

### Genomic feature learning

Several features from the genomic data are considered to be related to EPIs, such as the measures of open chromatin, DNA methylation, gene expression, ChIP-seq peaks for transcription factors, architectural proteins, and modified histones, and these features about the enhancers and promoters in our study can be directly obtained from Targetfinder [9]. Here, we use these features for each enhancer/promoter and the windows regions between them, denoted as $f_e$, $f_p$ and $f_w$, and take them into consideration to enhance the performance of prediction models.

For an enhancer-promoter pair, we consider several ways of fusing features of the enhancer and promoter $f_e$ and $f_p$, including subtraction, Hadamard product, and bilinear similarity. The subtraction is to consider the difference of vectors, denoted as $f_e - f_p$; Hadamard product is to consider local closeness between vectors, denoted as $f_e \odot f_p$; the bilinear similarity is calculated by $f_e^T W f_p$, where $W$ is a trainable weight matrix. Besides, we consider the genomic features $f_w$ of the window between the enhancer and promoter.

Finally, we obtain the representation of genomic features denoted as $h^G$ for the EPI prediction:

$$h^G = [f_e - f_p; f_e \odot f_p; f_e^T W f_p; f_w]$$

### EPI Prediction

For an enhancer-promoter pair, we combine the features from the sequences and genomic data, and obtain hybrid EPI features $h^{SG}$:

$$h^{SG} = [h^S; h^G]$$

Then the hybrid EPI features are feed into a multilayer perceptron (MLP):

$$y_{out} = \sigma(W h^{SG} + b)$$

The MLP has only one hidden layer with 128 nodes. We use Batch Normalization (BN) and dropout layer to prevent overfitting and accelerate the convergence of the network. $W$ is the trainable weight matrix, and $b$ is the bias. The sigmoid activation function $\sigma(.)$ is used to output the final predicted probability, which determines whether there is an interaction between the given promoter-enhancer pair.

### Model Optimization

EPIHC is implemented using Keras and is trained in an end-to-end manner. We use 6-mers for the sequence representation. We set the convolution parameter $s_{ck}$ as 40 and the number of kernels $n_f$ as 64, and use the RELU activation function. For max pooling layer, we set the window size $s_{mp}$ as 20 and the step size is also 20. Then, the dimension of communicative features $K$ is set as 128. For the model training, we set the batch size of the input data as 128, and the number of training epochs as 5, and adopt the cross-entropy loss function with L2 regularization. We set the learning rate as 3e-4 and use Adam optimizer.

## Results and Discussion

### Performance Evaluation

We adopt the five-fold cross-validation to evaluate the performances of models. As shown in Table 1, the benchmark datasets have the ratio of positives and negatives 1:20, and the training dataset in each fold of the cross-validation is also imbalanced, and previous studies[15-17] usually apply the data augmentation technique to the training datasets to alleviate the data imbalance. Specifically, for an enhancer-promoter pair, a window with the same length as the enhancer/promoter moves along their located chromosomes, and generate new pairs as new negatives. Therefore, the model is based on the augmented balanced training dataset and then is evaluated on the original testing set.

As discussed in [25, 26], the enhancer and promoter sequences from the same chromosomes have high redundant information and lead to the exaggerated performances of existing EPI prediction methods. Thus, we process the benchmark datasets in Table 1 to remove samples with high redundancy. For each cell line, we classify enhancer-promoter pairs in the datasets into 23 classes according to their located chromosomes. In the five-fold cross-validation, we classify them into 5 subsets. In each fold, we randomly select 4 subsets and use all samples in them as the training set, and then make predictions for the samples in the remaining subset (testing set).

In this way, the training set and testing set have no sample from the same chromosome. We name the above datasets as *chromosome-split datasets* in the following studies.

This study uses AUC and AUPR as indicators for model performance. AUC (Area Under Curve) refers to the area under the receiver operating characteristic curve (ROC). The ROC reflects the relationship between the sensitivity and specificity under the arbitrary classification threshold of the model. The closer the AUC value is to 1, the better the classification capacity of the model[23]. AUPR is the area under the Precision-Recall Curve, which focuses on the relationship between precision and recall, and it can well measure the performances of the model under imbalanced data[24]. Therefore, we use AUC and AUPR to comprehensively evaluate the prediction performances of models.

## Comparison with State-of-the-art methods

In this section, we evaluate the performances of the proposed method EPIHC, and compare it with three benchmark methods: SIMCNN[14], SPEID[13], and EPIVAN[16], which are based on the deep learning techniques. SIMCNN uses CNN to extract specific subsequence features or motifs for EPI prediction; SPEID uses CNN and RNN to capture long-term dependency in enhancer/promoter sequence; EPIVAN applies the attention mechanism after CNN and RNN to strengthen the representation of sequence information.

**Performance of methods on benchmark datasets**

First, all prediction models are evaluated on the benchmark datasets using five-fold cross-validation. As shown in Figure 2, EPIHC produces greater AUC scores than SIMCNN and SPEID for all cell line datasets, and produce greater AUC scores than EPIVAN for thee cell line datasets. When compared with EPIVAN, EPIHC produce better results on four cell line datasets GM12878, HUVEC, HeLa-S3 and K562, and similar results on others. The AUPR scores of EPIHC on all cell lines are much greater than that of SIMCNN, SPEID and EPIVAN, and EPIHC improves the AUPR scores of the best method EPIVAN by about 2.2%-3.6%.

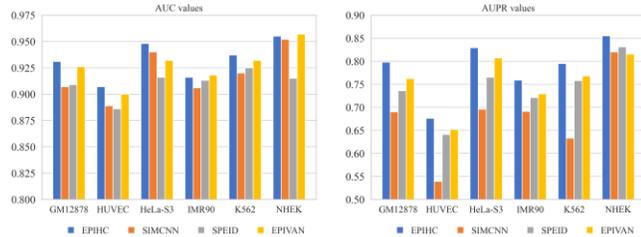

**Figure 2.** The performances of different methods on benchmark datasets.

Moreover, we use the communicative learning module in EPIHC, which has higher efficiency than existing deep learning-based EPI prediction methods. Therefore, we compare the training time of different models. Here, the training time includes the training time of all rounds and excludes the time for the data import and other irrelevant factors. As shown in Table 2, EPIHC needs much less time for model training, and costs about 4.5% of training SIMCNN, 2.3% of training SPEID, and 11% of training EPIVAN. As we know, SIMCNN is based on CNN with a great number of (about 300) convolution kernels, SPEID and EPIVAN depend on RNN to capture long-term dependency in enhancer/promoter sequence, but training RNN usually cost lots of time, while EPIHC utilizes the communicative learning module that has higher computational efficiency.

**Table 2.** Training time (seconds) of different models on benchmark datasets

| Model/Cell line | GM12878 | HUVEC | HeLa-S3 | IMR90 | K562 | NHEK |
|---|---|---|---|---|---|---|
| EPIHC | **831** | **608** | **695** | **499** | **787** | **513** |
| SIMCNN | 18288 | 13258 | 15061 | 11077 | 17562 | 11385 |
| SPEID | 36496 | 25864 | 29567 | 21149 | 33450 | 21621 |
| EPIVAN | 6694 | 6319 | 6450 | 4400 | 8396 | 4369 |

In general, EPIHC improves the predictive performances of existing EPI prediction models and also has a significantly better running efficiency.

**Performance of methods on the chromosome-split datasets**

The proposed method has satisfying performances on benchmarks, but the recent studies [25, 26] have revealed that the EPI prediction models are likely to produce the inflated performances on the benchmark datasets. Further, the proposed method and three benchmark methods are evaluated on the chromosome-split datasets using five-fold cross-validation.

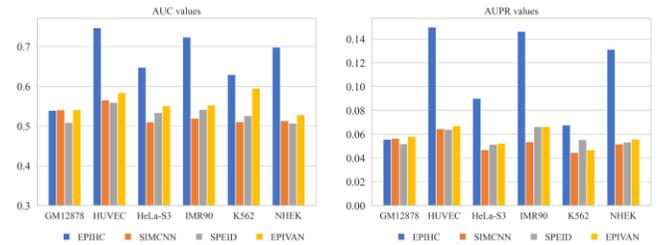

**Figure 3.** The performances of different methods on chromosome-split datasets.

As shown in Figure 3, the performances of all methods on chromosome-split datasets are much lower than the performances of methods on benchmark datasets, having about 18%-42% decrease in AUC scores and 85%-93% decrease in AUPR scores. We must emphasize that the experiments on chromosome-split datasets test the capability of EPI prediction models under extremely strict conditions. Nevertheless, EPIHC still produces better results than the compared methods for almost all datasets and has significant advantages. In general, EPIHC fully captures the potential pattern of EPIs, and demonstrate superior performances in the strict chromosome-split experiments.

## Discussion on EPIHC

The superiority of EPIHC has been well demonstrated by the above experiments, and the success of the EPIHC is owing to the design of EPIHC: hybrid features combination and the communicative learning module. The hybrid feature combination makes use of genomic features and sequence-derived features, and sequence-derived features include sequence-derived communicative features we describe above, and sequence-derived global features that are extracted by convolution and global max pooling. The communicative learning module considers the communication between segments of enhancer and promoter sequences to learn sequence-derived communicative features. Here, we consider the following variants to figure out the importance of hybrid features and the communicative learning module.

- **EPIHC-NSG** (NSG: no sequence-derived global features): we build the EPIHC model without sequence-derived global features.
- **EPIHC-NSC** (NSC: no sequence-derived communicative features): we remove the communicative learning module, and build the EPIHC model without sequence-derived communicative features.

- **EPIHC-NG** (NG: no genomic features): we build the EPIHC model without genomic features.

First of all, we compare the performance of EPIHC and three variants on the benchmark datasets, and the results are demonstrated in Figure 4. In general, all variants that use part of features lead to decreased performances, especially in terms of AUPR scores, and results show that all components are useful for EPIHC. The variants EPIHC-NSG and EPIHC-NSC have the greater drop in AUC and AUPR scores than EPIHC-NG, indicating that the sequence information has a significant impact on the performances of EPIHC, and that is why many sequence-based deep learning models have been proposed for the EPI prediction. EPIHC-NG and EPIHC have similar performances on benchmark datasets, and it seems that the genomic features make limited contributions to EPIHC.

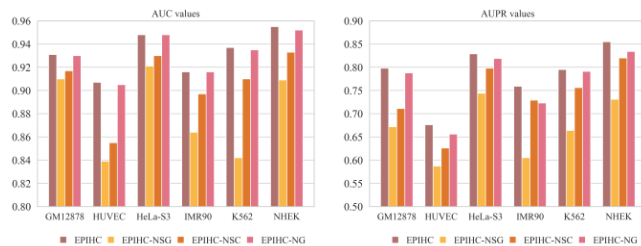

**Figure 4.** The performances of EPIHC and three variants on benchmark datasets

Further, we compare the performance of EPIHC and three variants on the chromosome-split datasets. As shown in Figure 5, similarly, all variants have lower performances than EPIHC. However, EPIHC-NSG and EPIHC-NSC that don't use sequence-derived features have subtle drop in performances; in contrast, EPIHC-NG that have don't use genomic features has the sharpest drop in performances, indicating that genomic features are of the most importance in the chromosome-split prediction. Comparison of performances of EPIHC-NG on benchmark datasets and chromosome-split datasets demonstrate that the genomic features also play the important role in EPI prediction when the training samples and testing samples are from different chromosomes. Although genomic features don't lead to very high performances, they provide important supplementary information that is free from the sequences.

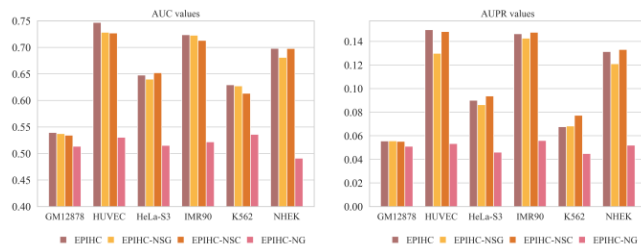

**Figure 5.** The performance of EPIHC and three variants on the chromosome-split datasets.

From enhancer and promoter sequences, we extract sequence-derived global features using CNN, and learn communicative features using the communicative learning module. Further, we pay attention to the communicative learning module, and thus compare the EPIHC models with and without the communicative learning module on the benchmark datasets, and investigate how it captures the sequence information ignored by CNN. The results show in Figure 6 demonstrate that the use of the communicative learning module greatly enhances the EPI prediction, gaining 1.5%-6% increase in AUC scores and 3.8%-12.2% increase in AUPR scores.

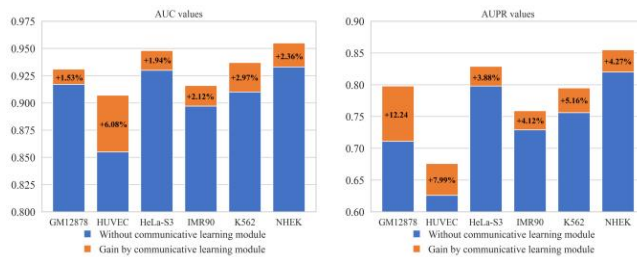

**Figure 6.** The performances of EPIHC w/o communicative learning module.

More importantly, the communicative learning module can provide interpretation about EPIs to some degree. The segment-level representation of the enhancer and the promoter for the communicative learning module are overlapped along the sequences. In the datasets, an enhancer has the length of 3000, and a promoter has the length of 2000, generating 148 vectors and 98 vectors as described in section *Feature extraction*. The communication map is calculated based on the representation of segments, and reveal the communication between segments. Here, we take the interaction between the enhancer chr9:4755000-4755644 and the promoter chr9:4983705-4986625 as an example, and visualize the communication map between the enhancer and the promoter. As shown in Figure 7, specified segments of enhancer/promoter have significant interactions with promoters/enhancers. For example. the 73-th contextual vectors of the enhancer and the 53-th contextual vectors of the promoters have the greatest communicative score, indicating that their communication plays the important role in the interaction between the enhancer chr9:4755000-4755644 and the promoter chr9:4983705-4986625.

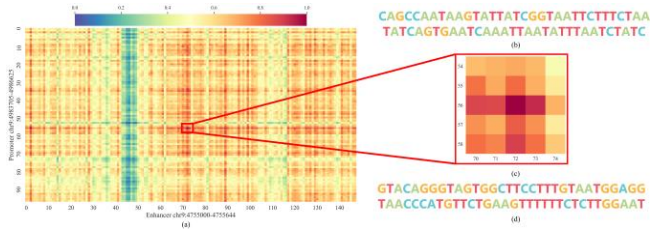

**Figure 7.** The communication map of enhancer (chr9:4755000-4755644) and promoter (chr9:4983705-4986625). (a) is an overview of the communication map based on the segment-level representations of enhancer and promoter. (b),(d) and (c) shows the details of 73-th contextual vectors of enhancer chr9:4755000-4755644 and 53-th contextual vectors of promoter chr9:4983705-4986625, including communication score and corresponding segments.

The studies demonstrate that the sequence-derived global features and sequence-derived communicative features greatly influence the performances of EPIHC, and the genomic features have an outstanding discriminative ability for the chromosome-split EPI prediction. Therefore, combining diverse features lead to the good performances of EPIHC.

### Strategies for improving cross-cell line prediction

The application of prediction models to new or unseen cell lines so-called *cross-cell line prediction* is very important. In machine learning, transfer learning is a technique where a model developed for a task is reused for another task. Utilizing the hidden patterns shared by different cell lines is the key to building the prediction models with great transfer capability. Since we have data from six cell lines, we try to build prediction models based on five cell lines and use them to predict the remaining one (target cell line).

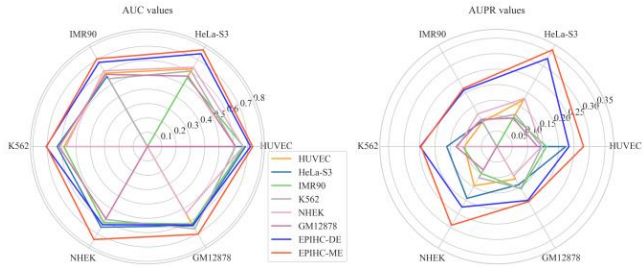

**Figure 8.** The performances of two improved models (EPIHC-DE and EPIHC-ME) and individual cell line-based models for testing cell lines

First, we build the prediction models based on one out of five cell lines and apply them to the target cell line. As shown in Figure 8, the cross-cell line prediction use the training set and the testing set come from different cell lines, the performance of EPIHC significantly decreases, the AUC value decreases by 30%-40%, and the AUPR value decreases by 50%-80%. As reported, the poor performance in cross-cell line prediction is because of the specificity of EPIs in different cell lines, and it is very important to enhance the performances of EPIHC in cross-cell line prediction.

Here, we propose two strategies for EPIHC to build prediction models, which have better transfer capability for cross-cell line prediction.

- **EPIHC-DE** (DE: data-ensemble strategy): the strategy is to directly combine the datasets from all available cell lines, and then build the prediction model on the combined datasets, and then apply it to the target cell line.
- **EPIHC-ME** (ME: model-ensemble strategy): the strategy is to consider the differences of available cell-lines, and build the individual sub-models based on each cell-lines, and combine them with the attention mechanism, and then apply it to the target cell line.

The results of EPIHC-DE and EPIHC-ME in cross-cell line prediction are shown in Figure 8. Since both ensemble strategies take advantage of common information of EPIs in different cell lines, EPIHC-DE and EPIHC-ME have the significant improvement on AUC and AUPR scores when compared to EPIHC. Compared with EPIHC-DE, EPIHC-ME produces slightly better performances, because EPIHC-ME differently treats data from different cell lines.

Further, we analyze the attention weights of individual cell lines used by EPIHC-ME, which reflect the contributions of individual available cell lines to the prediction on the target cell line. Since we have six cell lines, we use a cell line as the target, and build the model based on the other five cell lines, and apply it to the target line. Since the weights learned by the attention mechanism is varied for different testing samples, weights for all testing samples in a target cell line are analyzed statistically. As shown in Figure 6, EPIHC-ME captures the prominent information of available cell lines most related to unseen cell line. For example, data from the cell line HeLa-S3 have greater weights than others when evaluated on the target cell line K562; the cell line HUVEC-related attention weights has obvious greater contributions than others when evaluated on the target NHEK.

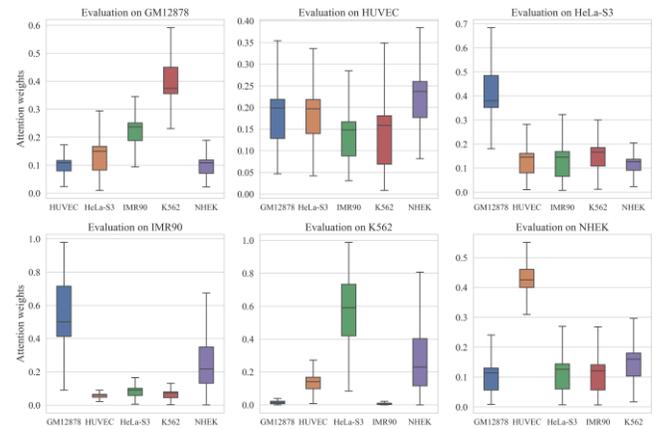

**Figure 9.** Attention weights of data sources used in EPIHC-ME

The cross-cell line prediction is indeed a challenge, and we take two ensemble strategies for EPIHC to build the prediction models, which demonstrate good transfer capability in the computational experiments.

## Conclusions

This study focuses on the prediction of the interactions between enhancers and promoters (EPIs). EPI prediction is extremely important for understanding the genomic processes in cell lines. We propose an EPI prediction method based on deep learning by integrating the sequence data and genomic data of enhancers and promoters. We introduce a communicative learning module to capture the communicative information between enhancers and promoters, which were not considered in previous deep learning-based EPI prediction models. Moreover, the communicative information learned by the communicative learning module, the global sequence information and the genomic features provide hybrid EPI information for building high-accuracy prediction models. The experiments demonstrate EPIHC produces better results than existing state-of-the-art methods on benchmark datasets and chromosome-split datasets, and the ablation analysis figures out the contributions of each component of EPIHC. At last, we pay attention to the problem of how to enhance the cross-cell line prediction, and propose two strategies that can further improve EPIHC.

---

**Key Points**

- We design a communicative learning module for EPIHC, which captures segment-level communicative information between enhancer and promoter sequences as well as the sequence dependency. More importantly, the module has high efficiency.
- EPIHC utilizes hybrid features, including both sequence-derived features and genomic features, and the combination of diverse information can lead to high-accuracy prediction models. The studies revealed that both sequence-based features and genomic features are very important for the EPI prediction.
- Two strategies including data-ensemble and model-ensemble are designed for EPIHC to enhance the cross-cell line prediction. Experimental results show that models trained on available cross-cell lines perform well on other cell lines

---

## Supplementary data

No supplementary.

**Funding**

This work was supported by the National Natural Science Foundation of China (Grant No. 62072206, Grant No. 61772381, Grant No. 61572368), Huazhong Agricultural University Scientific & Technological Self-innovation Foundation. The funders have no role in study design, data collection, data analysis, data interpretation, or writing of the manuscript.